\documentclass[aps,prb,twocolumn,superscriptaddress,showpacs,longbibliography,nofootinbib]{revtex4-2}
\usepackage{amsmath}
\usepackage{graphicx}
\usepackage{amsfonts}
\usepackage{verbatim}
\usepackage{amssymb}
\usepackage{color}
\usepackage{dsfont}
\usepackage{amsmath} 
\usepackage{hyperref}
\usepackage{physics}
\usepackage{soul}
\hypersetup{colorlinks = true, urlcolor = blue, linkcolor = blue, citecolor = blue}
\usepackage[normalem]{ulem}

\newcommand{\bpm}{\begin{pmatrix}}
\newcommand{\epm}{\end{pmatrix}}

\newcommand{\expect}[1]{\langle  #1\rangle}

\newcommand{\cmtc}{Condensed Matter Theory Center and Joint Quantum Institute, Department of Physics, University of Maryland, College Park, Maryland 20742-4111, USA}

\begin{document}

\title{Capacitance-based Fermion parity read-out and predicted Rabi oscillations in a Majorana nanowire}
\author{ Jay D. Sau}
\affiliation{\cmtc}
\author{Sankar Das Sarma}
\affiliation{\cmtc}
\date{\today}

\begin{abstract}
Recent experiments have measured flux dependent capacitance at radio frequencies leading to the potential for a fast parity 
read-out of a Majorana qubit. In this work we argue that the quantum dot used in the capacitance measurement can be reasonably 
approximated by a non-interacting weakly coupled orbital. We then predict the measured flux and parity 
dependent capacitance for several parameter regimes of the disordered Majorana nanowire model that are both topological and trivial. 
Following this we study how such a fast capacitance read-out can be used to characterize the quantum coherence of a Majorana 
nanowire-based qubit using Rabi oscillations. We additionally show that such measurements, if made possible by  coherent inter-wire tunneling,
would provide a valuable way of characterizing the low energy states in the frequency domain.
\end{abstract}

\maketitle

\section{Introduction}\label{S1}
\begin{figure}[t]
\begin{center}
\includegraphics[width=0.46\textwidth]{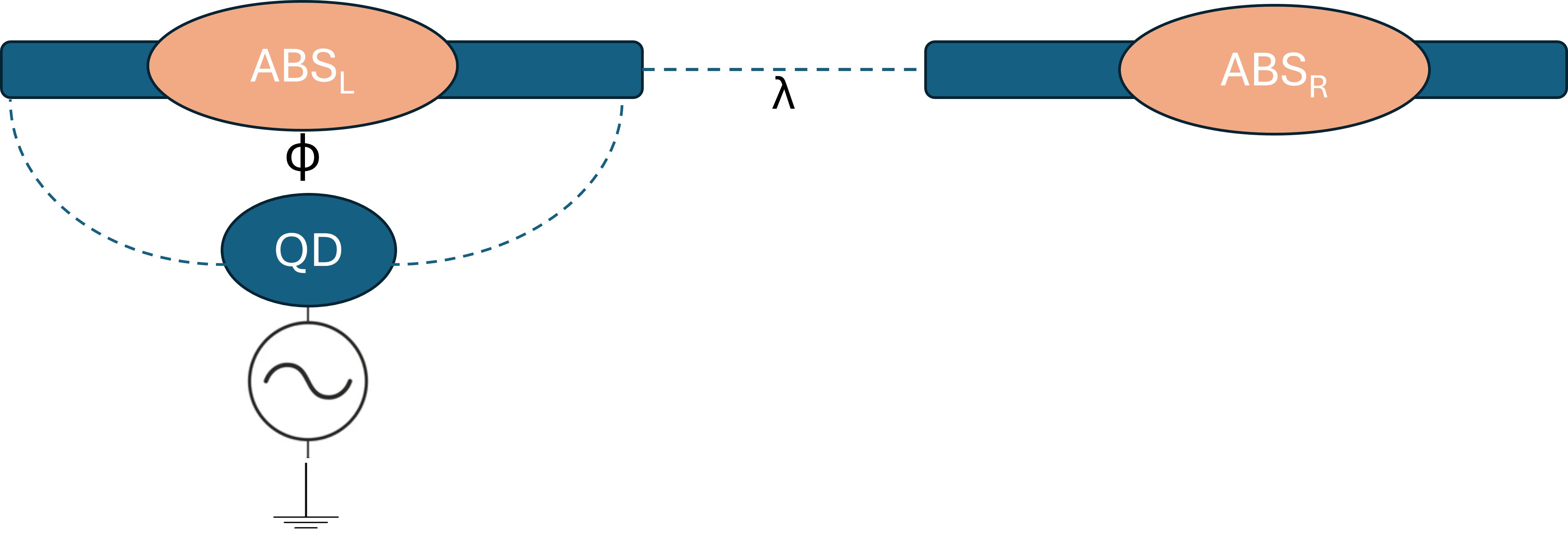}
\end{center}
\vspace{-3mm}
\caption{ Schematic for Majorana nanowire device, similar to recent experiment~\cite{aghaee2024interferometric}, containing  a pair of Majorana nanowires to measure Rabi oscillation. Both Majorana nanowires are assumed to be tuned into a parameter regime to support a low energy ABS, ideally formed from the pair of MBSs in each wire. The qubit is formed 
from the occupancy of the occupancy of $ABS_{L,R}$, which are expected to be at zero energy in the absence of tunneling (i.e. $\lambda=0$).
The Majorana wire on the left containing $ABS_L$ is attached to a QD at both ends, forming a loop that encloses magnetic flux $\Phi$ that is 
controlled by a small perpendicular field. Measuring the flux dependent capacitance of the QD allows a fast read-out of the occupancy of $ABS_L$. Turning on tunneling $\lambda\neq 0$ splits the $ABS_{L,R}$ leading to Rabi oscillations in the occupancy of $ABS_L$ as a function of time that can establish the coherence of MBS qubits.}
\label{FIG1}
\vspace{-1mm}
\end{figure}

The topologically protected degree of freedom associated with a Majorana system is the Fermion parity of the pair of Majoranas at 
the end of a wire. This Fermion parity is equivalent to the occupancy of a low-energy Andreev bound state (ABS), seen in Fig.~\ref{FIG1}, 
that is formed from the pair of Majorana zero modes (MZMs).  The addition of tunneling between the ends results in a closed loop~\cite{Kitaev2001Unpaired} with the ABS moved to finite energy, which is shown on the left half of Fig.~\ref{FIG1}.
In the case of an ideal topological wire, the quantum 
information (i.e. Fermion parity) of the pair of Majorana modes transforms into the occupancy 
of this finite energy bound state. Any scheme to measure the occupancy of this finite energy bound state  by measuring 
inductance~\cite{Lutchyn2010Majorana}, conductance~\cite{Sau2015Proposal,Liu2019Proposal} or by a capacitance measurement~\cite{Plugge2017Majorana,Karzig2017Scalable} constitutes a 
read-out of the Majorana qubit. Such a capacitance measurement-based parity readout in a Majorana nanowire~\cite{Lutchyn2010Majorana,Oreg2010Helical}
has recently been accomplished using a fast RF circuit~\cite{aghaee2024interferometric}. 
A capacitance measurement contains contributions other than that from the occupied bound state. Therefore only 
the change in the capacitance resulting from a $\Phi_0$ change in the magnetic flux $\Phi$ through the loop shown in Fig.~\ref{FIG1} 
is used to determine the bound state occupancy. This interferometric flux dependence of the capacitance is related to the fractional Josephson effect~\cite{Kwon2004Fractional,Lutchyn2010Majorana} as well as non-local teleportation~\cite{Fu2010Electron}.

However, such an interferometric measurement of the bound state occupancy is affected by non-idealities such as the short system size 
relative to the effective coherence length~\cite{Liu2019Proposal}. This can be a problem even in nominally long wires such as the $3\,\mu m$ long wires used 
in recent experiments~\cite{aghaee2024interferometric} because 
of sub-gap states generated by disorder~\cite{das2023spectral,das2023density}. 
This makes it critical to investigate such a capacitance measurement using a general model of a disordered Majorana nanowire, where
one is not limited to the low-energy subspace of the lowest Andreev states in the wire~\cite{aghaee2024interferometric}. Given the demonstration 
of  fast parity read-out~\cite{aghaee2024interferometric} in a class of cleaner devices that have shown improved potential to realize MZMs, one may 
hope to demonstrate a higher level of coherence  than was obtained in past Majorana nanowire devices~\cite{vanZanten2020Photon}, 
thus paving the way toward a Majorana qubit.

In this work, we compute the ac capacitance of a Josephson junction formed by connecting the ends of a disordered Majorana nanowire 
shown in Fig.~\ref{FIG1}. The capacitance is equivalent to the response of the charge in the quantum dot (QD) in Fig.~\ref{FIG1} in response
to modulating the potential of the QD. This is calculated using linear response of the Josephson junction formed through the QD, using a formalism that is similar to the calculation of the Josephson inductance~\cite{tewari2012probing}. We study this response for the disordered 
Majorana nanowire model in a variety of regimes including the Andreev bound state (ABS) and topological patch regime~\cite{das2023spectral,das2023density}. 
The fast parity read out can be used to measure Rabi oscillations. 
We then compute the coherence, frequency and amplitude of the Rabi oscillation in topological qubits created by tunnel coupling pairs of Majorana wires in the corresponding parameter regimes.

\section{Parity read-out}\label{S2}
The left Majorana nanowire supporting $ABS_L$ shown in Fig.~\ref{FIG1} contains a QD tunnel coupled to both ends of the Majorana nanowire 
with a flux $\Phi$ enclosed by the loop. The essential function of the QD in the experiment~\cite{aghaee2024interferometric} is to provide 
tunneling between the ends of the nanowire through elastic co-tunneling~\cite{aleiner2002quantum}, 
 which includes cross-Andreev processes that maybe comparable to the direct tunneling~\cite{bolech2007observing,tewari2008testable,Fu2010Electron}.
 The Coulomb interaction in the QD is essential to enhance the single particle (charge) gap of approximately $45\,\mu eV$~\cite{aghaee2024interferometric} in the QD beyond the single-particle level spacing of $20\,\mu eV$, which would be quite low for a $2.4\,\mu$m long QD~\cite{aghaee2024interferometric}. For weak tunneling that leaves the QD in the strongly Coulomb blockaded regimes~\cite{aghaee2024interferometric}, we can represent the states in terms of a single electron creation operator $d^\dagger$ corresponding to the single-particle level closest to the Fermi level  together with a host of lower energy particle-hole excitations in the QD levels~\cite{aleiner2002quantum}. For sufficiently low voltages, frequencies and tunnel coupling, these particle-hole excitations are gapped by the single-particle level spacing so that their contribution to the co-tunneling process is a perturbative renormalization of the tunneling amplitude. In this limit, the coherent elastic co-tunneling across the QD is phenomenologically identical to a tunneling amplitude between the ends of the nanowire. Furthermore, the elastic co-tunneling is modulated by the gate voltage $V_{QD}$ at the QD that is also used to measure the capacitance. We model the low-energy properties of the QD at a phenomenological level by tunneling through a single non-interacting orbital with an energy level at $V_{QD}$. The magnitude of the tunneling is adjusted to match the measured value of capacitance~\cite{aghaee2024interferometric}. Potential complications of the multi-level interacting QD physics that not captured by this renormalized parameter model will be discussed in the conclusion.
 
 The model for the QD described above allows us to understand the parity measurement using the configuration of the left Majorana nanowire in Fig.~\ref{FIG1} in terms of a non-interacting Bogoliubov de Gennes (BdG) Hamiltonian, $H(\phi)$,
that is the Majorana nanowire model~\cite{DasSarma2016How} (see Appendix.~\ref{SecA1} for details).
The magnetic flux $\phi$ in the measuring loop containing the QD in 
 Fig.~\ref{FIG1} is in units where the periodicity of the Hamiltonian in $\phi$ i.e. the flux quantum is set to $hc/e=2$.
 We quantify the degree to which the wire is topological by the topological stability $T_S$ defined  as the product of the class D topological invariant~\cite{Kitaev2001Unpaired} with the ratio of lowest eigenvalues for the Hamiltonian $H$ (without the QD) with flux $\phi=0$ and $\phi=1$(see Appendix~\ref{SecA1} for details). This metric, apart from measuring the topological invariant (indicated by a negative sign), can indicate a short wire when this ratio differs substantially from unity. Thus a value of $T_S$ substantially larger than -1 indicates a poorly defined topological phasse because of finite size effects~\cite{das2023spectral}.  The hopping efficiency through the QD (for our numerical results) is chosen to be $\lambda_{QD}=0.03$ so as to obtain energy splittings in the range of several tens of mK (consistent with the current experimental situation).
When studying disordered wires, we choose the disorder potential for the Majorana nanowire to be Gaussian distributed with a correlation length of $\xi_{dis}\sim 30 nm$ with a root mean square amplitude $V_{RMS}$. 

The  quantum capacitance measured from this gate (up to a lever arm) is given by $C=-d^2 E_{tot}/dV_{QD}^2$, where $E_{tot}=\sum_{E_m<0}E_m$ 
is the total energy of the system. The finite frequency generalization of this quantum capacitance can be obtained for the above nanowire Hamiltonian using linear response theory as:
\begin{align}
&C(\omega)=\Gamma_C\sum_{E_n<0,E_m>0}\frac{|\expect{E_n|\tau_z\delta_{r,0}|E_m}|^2(E_n-E_m)}{(E_n-E_m)^2-\omega^2},\label{eq:Capacitance}
\end{align}
 where $\Gamma_C=2\alpha^2 e^2/k_B=1.9\, fF-K$ is a constant that converts the capacitance $C(\omega)$ to femtofarads (fF) when the energies $E_n$ are in units of Kelvin (K) after adding a lever arm $\alpha=2^{-1/2}$
 that would be relevant to recent experiment~\cite{aghaee2024interferometric}. 
In the above $\ket{E_n}$ are eigenstates of the BdG Hamiltonian of the system, and $\tau_z$ is the matrix coupling to charge density, both written in the Nambu basis (see Appendix~\ref{SecA1} for details).
 The read-out scheme used in the recent experiments~\cite{aghaee2024interferometric} involves studying the dependence of $C$ on the magnetic flux $\phi$ as well as the 
 parity of the wire, which can be changed by switching the occupancy of the state of the smallest energy states from $0$ to $1$ and vice-versa. 

\begin{figure}[t]
\begin{center}
\includegraphics[width=0.48\textwidth]{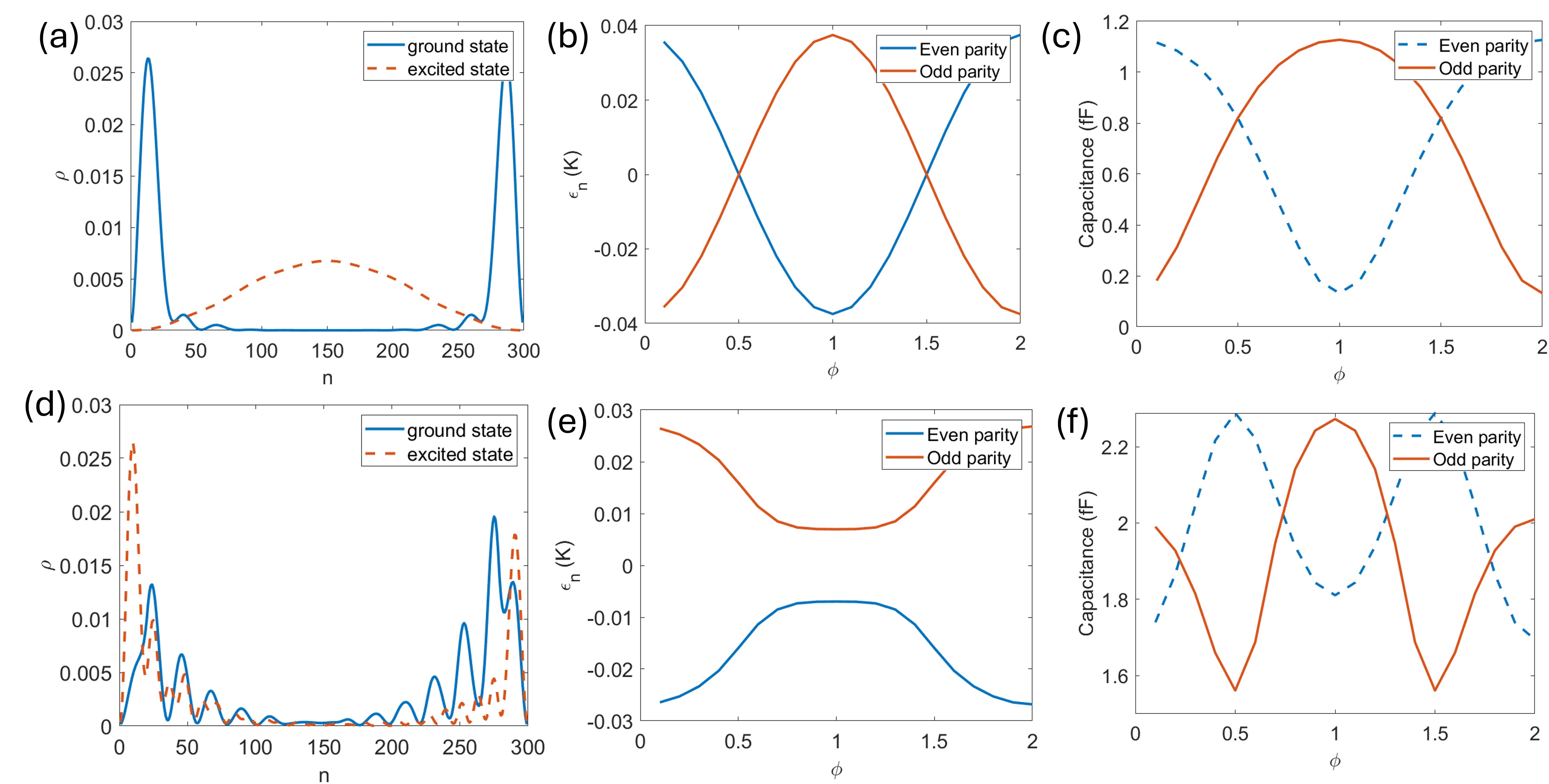}
\end{center}
\vspace{-3mm}
\caption{(a) Wave-function of the lowest energy eigenstate, (b) energy in K and (c) capacitance for the occupied and unoccupied state as a function of magnetic flux $\phi$ (in units where $hc/e=2$) for an $L=3\,\mu m$  Majorana nanowire without disorder at $\mu=0\,K$, $B_x=0.28\, T$ and $V_{QD}=0.5\,K$. The wave-functions for this ideal case show clear presence of end MZMs with a topological stability~\cite{Kitaev2001Unpaired} (see Appendix~\ref{SecA1} for details) $T_S=-0.99$ and MZM splitting for the isolated wire $E_M=30\,\mu K$.  As expected~\cite{aghaee2024interferometric} we see a shift of $\Delta\phi=1$ between the even and odd parity sectors. The panels (d,e,f) are the corresponding result for the Majorana nanowire in a non-topological phase with $T_S=0.49$. The parameters for the phase are $B_x=0.4\, T$, $\mu=5.5\, K$, $V_{QD}=5.5\, K$ and $L=3\,\mu m$. The low energy states are ABSs generated by a potential (see text for details) with parameters $V_{abs}=4.5\, K$ and $L_{abs}=200\, nm$. }
\label{FIG2}
\vspace{-1mm}
\end{figure}

\subsection{Parity results}\label{S2a}
Consistent with the results of the  experiment~\cite{aghaee2024interferometric}, we find that the capacitance read-out for an 
ideal Majorana nanowire shows an $hc/2e$ shift in flux between even and odd parity states. Fig.~\ref{FIG2}(a) shows the lowest energy (positive) state, 
which is composed of a pair of MZM wave-functions 
that are localized at the end for the optimal parameters (see caption) of a clean nanowire  with the next excited state being delocalized.
The individual MZM wave-functions can be constructed by considering superpositions of the lowest positive energy state, referred to as ground 
state in Fig.~\ref{FIG2}, with its particle-hole conjugate state with negative energy.
Introducing tunneling between the ends of the 
nanowire through a QD to form a loop with enclosed flux splits the end MZMs leading to a pair of flux dependent energy levels shown in Fig.~\ref{FIG2}(b). The two levels represent the energies of different Fermion parity states of the nanowire. It is clear that these levels 
interchange energies when the flux changes by $\delta \Phi=hc/2e$ as happens in the fractional Josephson effect~\cite{Kitaev2001Unpaired,Kwon2004Fractional,Sau2015Proposal}. Fig.~\ref{FIG2}(c) shows the corresponding capacitance, which 
reflects the response of each energy state to a variation in the tunneling strength which is modulated by the  potential of the QD. The
capacitance shows a similar characteristic to that of the energy. The flux and parity dependence of the predicted capacitance allows, 
in the ideal case, a read-out of the Fermion parity of the Majorana nanowire through essentially determining the level occupancy of the 
split MZMs.

While capacitance-based parity readout works well for ideal Majorana nanowires, its effectiveness in regimes~\cite{PanPhysRevResearch.2.013377}  with trivial zero-bias peaks from disorder or end potentials remains uncertain. In Fig.~\ref{FIG2}(d–f), we examine cases where zero modes originate from non-topological near-zero-energy ABSs, often induced by end quantum dot potentials~\cite{Kells2012Near,Prada2012Transport,Rainis2013Towards,Liu2017Andreev,Vuik2019Reproducing}. Fig.~\ref{FIG2}(d) shows such ABSs forming due to an end quantum well potential (see Appendix~\ref{SecA1}). Despite their low energy, Fig.~\ref{FIG2}(e) indicates that ABS spectra exhibit flux modulation smaller than the average energy, confirming the wire's non-topological nature. Capacitance readout in Fig.~\ref{FIG2}(e) reveals a clear parity difference, but the flux dependence is dominated by an 
$hc/2e$ periodic modulation that flips sign between parities, unlike the ideal Majorana case where the 
$hc/e$ component maintains its sign. This aligns with recent capacitance analyses based on a minimal low-energy ABS model~\cite{aghaee2024interferometric}.

\begin{figure}[t]
\begin{center}
\includegraphics[width=0.48\textwidth]{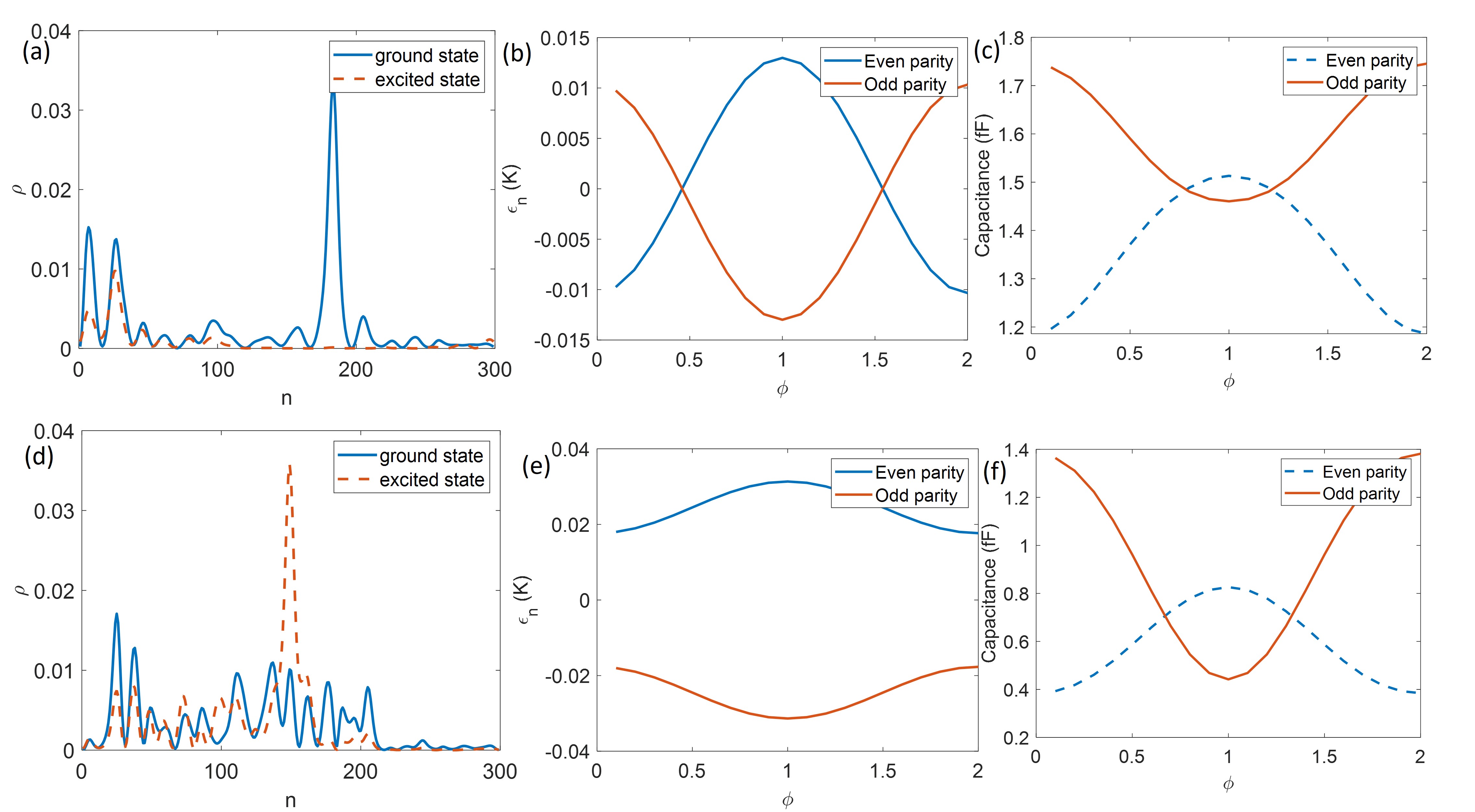}
\end{center}
\vspace{-3mm}
\caption{(a) Wave-function of the lowest energy eigenstate, (b) energy in K and (c) capacitance for the occupied and unoccupied state as a function of magnetic flux $\phi$ (in units where $hc/e=2$) for an $L=3\,\mu m$  Majorana nanowire in the topological patch regime~\cite{das2023spectral} at $\mu=0.2\,K$ and $B_x=0.42\, T$, $V_{QD}=0.2\,K$, disorder strength $V_{RMS}=7.5\, K$. The wave-functions for this case show a single low energy ABS with a topological stability~\cite{Kitaev2001Unpaired} (see Appendix~\ref{SecA1} for details)  $T_S=-0.621$. As expected~\cite{aghaee2024interferometric} we see an approximate shift of $\Delta\phi=1$ between the even and odd parity sectors. The panels (d,e,f) are the corresponding result for the Majorana nanowire in a non-topological phase with $T_S=0.13$. The parameters for the phase are $B_x=0.56\, T$, $\mu=2.0\, K$, $V_{QD}=2.0\, K$ $V_{RMS}=15\, K$ and $L=3\,\mu m$.}
\label{FIG25}
\vspace{-1mm}
\end{figure}

Note that while the detailed results in Fig.~\ref{FIG2} depend on a specific choice of parameters listed in the caption that are used in the Hamiltonian described in Appendix~\ref{SecA1}, the qualitative features are generic for the corresponding state i.e. topological MZM or ABSs. The parameters listed in the caption are in the range that experiments~\cite{aghaee2024interferometric} tune within and in this case are fine-tuned to produce a small MZM splitting or ABS end states. The parameter choice is motivated from previous descriptions of the phase diagram of the MZM nanowire model~\cite{Lutchyn2010Majorana,Oreg2010Helical} in the field~\cite{das2023spectral,Liu2017Andreev} among others and therefore will not be repeated here. 
The local zero-energy ABS model described above can practically be ruled out by three terminal measurements of the tunnel barrier 
dependence of the local and non-local conductance~\cite{aghaee2022inas}. However, fine-tuned low energy states resulting from 
intermediate strength disorder  called topological patches~\cite{das2023spectral,das2023density} can be a viable explanation 
for the topological regions seen experimentally~\cite{aghaee2022inas}. Fig.~\ref{FIG25}(a) shows the wave-function of the lowest energy eigenstate of the weakly disordered Majorana nanowire with parameters chosen to generate such a zero-energy state  following previous work~\cite{das2023spectral} on the model. Despite the topological stability with $T_S$ being 
 negative, which would nominally be defined as topological based on the usual topological invariant~\cite{Kitaev2001Unpaired}, 
the lowest energy ABS wave-function has rather small weight at the end and cannot reasonably be construed as being composed of 
end-localized MZMs. Despite this, Fig.~\ref{FIG25}(b) shows that the lowest energy state has a flux dependence with a substantial $hc/2e$ shift between different parity sectors, which is rather similar to the topological MZM spectrum seen in Fig.~\ref{FIG2}(b). Similar to the topological case, the capacitance in the topological patch regime seen in Fig.~\ref{FIG25}(c) shows a significant $hc/e$ periodic component.  Furthermore, this component flips sign 
similar to the topological case (i.e. Ref.~\ref{FIG2}(c)). However, unlike the ideal case, the amplitude of oscillations in the different parity sector are different as is seen in some of periods of the capacitance data measured in recent experiments~\cite{aghaee2024interferometric}. It should be noted that identifying the oscillation amplitude in the experiment is difficult because of the Aharonov Bohm oscillation amplitude varying between different periods of the oscillation due to magnetic field induced shifts.
Topological stability in the patch regime~\cite{das2023spectral} is highly disorder-dependent. Stronger disorder (Fig.~\ref{FIG25}(d)) leads to shorter localization compared to Fig.~\ref{FIG25}(a). Despite a trivial $T_S$ ($T_S>0$), Fig.~\ref{FIG25}(e–f) still shows substantial 
$hc/e$ components, resembling the topological case in the upper panels.
So that the capactiance based parity signal is not a definitive indicator for a topological phase.

\subsection{Constraints on topology from parity oscillations}\label{S2b}
Comparing the results for our simulated flux and Fermion parity dependence of the capacitance for the ideal MZM case (i.e.  Fig.~\ref{FIG2}(a,b,c))  to the other cases suggest that the
capacitance in the topological case is characterized by an 
$hc/e-$periodic flux dependence together with an exact $hc/2e$ phase shift between the different parity sectors.
Indeed, this signature is equivalent to both the fractional Josephson effect~\cite{Kwon2004Fractional,Kitaev2001Unpaired,Lutchyn2010Majorana,Sau2015Proposal} as well as teleportation~\cite{semenoff2006teleportationmajoranamedium,tewari2008testable,Fu2010Electron,Sau2015Proposal,Vijay2016Teleportation} and thus can place complementary constraints on the topological character of a system compared to transport. 

The fractional Josephson effect~\cite{Kwon2004Fractional,Kitaev2001Unpaired,Lutchyn2010Majorana,Sau2015Proposal} in the topological superconducting phase can be used to argue for a characteristic flux dependence of the capacitance. To understand this, note that for a "long" superconducting wire, which is longer than the coherence length, the introduction of a phase slip in the middle of the superconducting wire is equivalent to an $hc/2e$ phase shift. In the trivial phase with a large length, this cannot have an effect on the capacitance, which must then be $hc/2e-$ periodic. In contrast, in the topologically non-trivial phase, such a phase slip must be accompanied by a change in fermion parity~\cite{Kitaev2001Unpaired}. Thus, the capacitance in the topological case must be characterized by an $hc/2e-$ phase shift between the two fermion parity sectors, which must otherwise be $hc/e-$periodic. This argument assumes that the superconductor is infinitely long either in the topological or trivial phase. These conclusions are modified for a practical device where finite-size corrections, often strong need to be accounted for.
A gapped trivial phase shows an $hc/2e-$periodic flux dependence similar to Fig.~\ref{FIG2}(d,e,f). 
The topological patch regime Fig.~\ref{FIG25}, while showing a near $hc/2e$ phase shift between the different parity sectors, also shows a change in average as well as amplitude that are not allowed in the topological phase. The recent capacitance measurements in the Majorana nanowire~\cite{aghaee2024interferometric}, do not distinguish between MZM and the topological patch regime because of an ambiguous flux periodicity.

\begin{figure}[t]
\begin{center}
\includegraphics[width=0.46\textwidth]{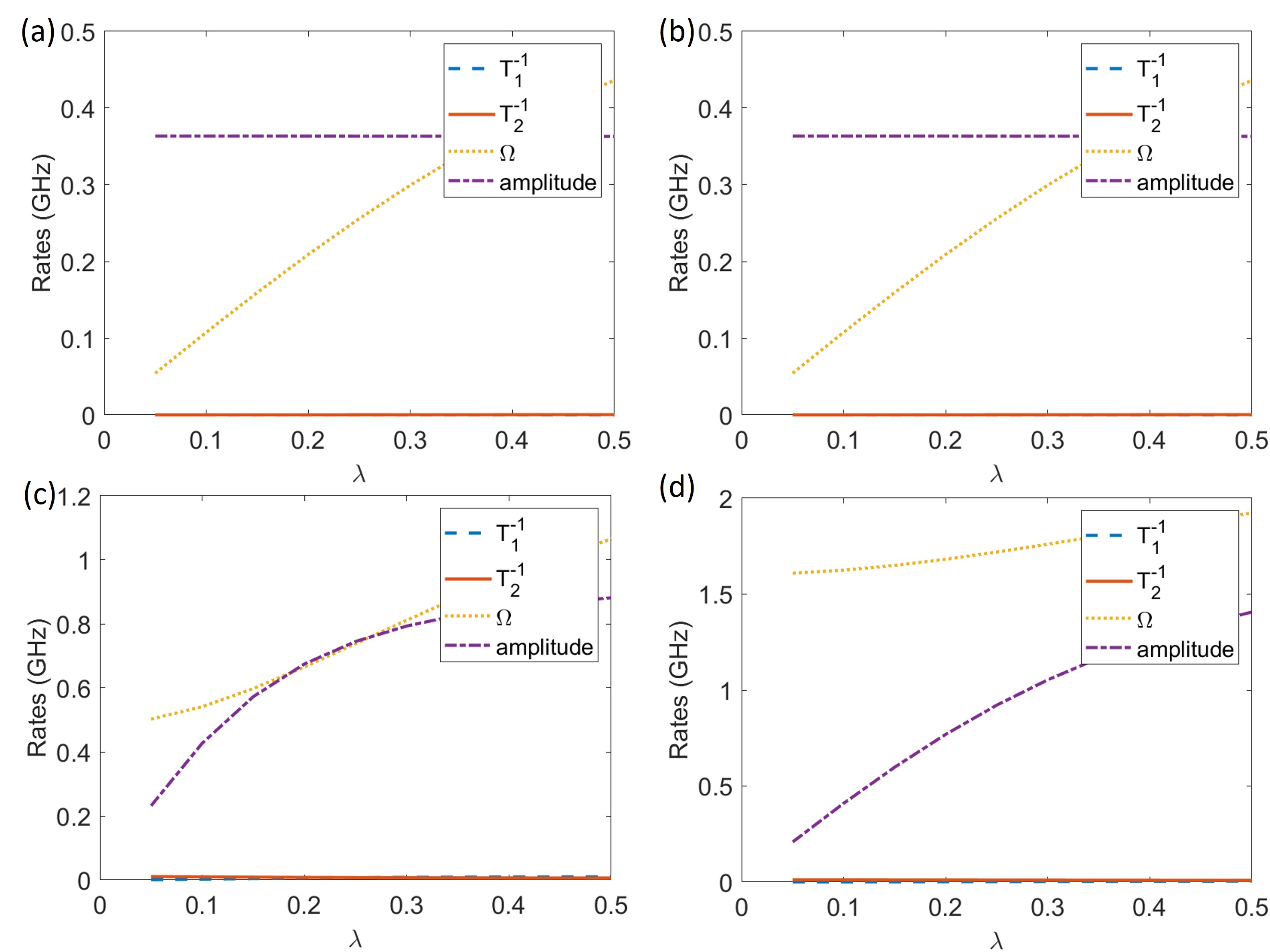}
\end{center}
\vspace{-3mm}
\caption{Rabi oscillation parameters relaxation rate $T_1^{-1}$, dephasing rate $T_2^{-1}$, Rabi-oscillation frequency $\Omega$ and amplitude (dimensionless with a scaling factor of $0.76$ for the top panels and $0.01$ panel (c) and $0.3$ for panel (d)) of Fermion parity oscillation as a function of 
inter-wire coupling $\lambda$. The left panels (right panels) refer to the case for even (odd) total Fermion parity respectively of the system shown in Fig.~\ref{FIG1}. The top panels (a,b) correspond to parameters of Fig.~\ref{FIG2}(a,b,c) with inter-wire parity decoherence rate $(T_1')^{-1}\sim 38\,Hz$ while the bottom one corresponds to parameters Fig.~\ref{FIG2}(d,e,f) with $(T_1')^{-1}\sim 0.01\,GHz$. }
\label{FIG3}
\vspace{-1mm}
\end{figure}

The above argument, while suggesting the necessity of a specific flux-periodicity for the topological superconducting phase, does not show the flux periodicity to be a 
sufficient condition for topology. To understand this, let us consider a finite length superconducting nanowire with a single low-energy ABS that has weight at both 
ends of the nanowire. A topological superconducting nanowire with a pair of MZMs at the ends would be an example of such a state, as would the topological patch regime 
described above. To be more specific we can represent this 
 low energy ABS or pair of MZMs in a wire 
by a fermion operator $c=(\hat\gamma_R+i\hat\gamma_L)2^{-1/2}$. Let us 
represent the low energy fermion operator on the site $N=0$ 
by $d$. Introducing the flux dependent phase $\phi$ (in units so that $hc/2e=1$) on the right tunnel junction: 
\begin{align}
    &H=\delta c^\dagger c+2^{1/2} t_R  e^{i\pi\phi}(\gamma_R+\zeta_R\gamma_L)d\nonumber\\
    &+2^{-1/2}t_L d(\gamma_L+\zeta_L\gamma_R)+h.c.,
\end{align}
where $t_{L,R}>0$ are hopping amplitudes through the left and the 
right ends of the wire, $\zeta_{L,R}$ represent the leakage amplitudes of the MZMs across the wire and $\delta$  is the MZM splitting. Writing the Hamiltonian in terms of the fermion operator c, we get
\begin{align}
    &H=\delta c^\dagger c +d[t_R e^{i\pi\phi}\{(1+i\zeta_R)c^\dagger +(1-i\zeta_R)c\}\nonumber\\
    &+t_L\{(i+\zeta_L)c^\dagger+(-i+\zeta_L)c\}]+h.c.
\end{align}
Note that choosing $\zeta_{R,L}$ to be complex allows $H$ to 
contain arbitrary complex normal and Andreev tunneling amplitude at each junction. The above Hamiltonian results from a projection of the BdG Hamiltonian in Eq.~\ref{eq:HParity} to the lowest energy states in the nanowire and the QD. The operator $c$ corresponds to the lowest pair of states in the Majorana nanowire and the operator $d$ corresponds to the electron in the quantum dot. The parameter $\delta$ is directly controlled by the splitting energy of the MZMs or the voltage $V_{QD}$ of the quantum dot. The parameters $t_{L,R}$ and $\zeta_{L,R}$ are matrix elements of the tunneling term (proportional to $\lambda_{QD}$) in Eq.~\ref{eq:HParity}. Since the conclusions of this subsection depend on whether the relative order of magnitudes of these parameters, we do not compute them for the specific results in Fig.~\ref{FIG2} and Fig.~\ref{FIG3}.

The two terms separate naturally by fermion parity sector.
In the even fermion parity sector $c^\dagger c+d^\dagger d=0,2$, the $cd^\dagger=c^\dagger d=0$ so that writing $d^\dagger d=c^\dagger c=(1+\nu_z)/2$ and $c^\dagger d^\dagger=\nu_+$ we get 
\begin{align}
    &H_{even}(\phi)=\delta \nu_z/2\nonumber\\
    &+\{t_R e^{i\pi\phi}(1-i\zeta_R)+t_L(-i+\zeta_L)\}\nu_-+h.c.
\end{align}
Here $\nu_{x,y,z}$ are $2\times2$ Pauli matrices representing the fermion parity of the nanowire Fermion state $c^\dagger$. The matrices $\nu_\pm=(\nu_x\pm i\nu_y)/2$.
In the odd fermion parity sector 
 $c^\dagger c+d^\dagger d=1$, the $c^\dagger d^\dagger=c d=0$ so that writing $1-d^\dagger d=c^\dagger c=(1+\nu_z)/2$ and $d c^\dagger =\nu_+$ we get 
\begin{align}
    &H_{odd}(\phi)=\delta \nu_z/2\nonumber\\
    &+\{t_R e^{i\pi\phi}(1+i\zeta_R)+t_L(i+\zeta_L)\}\nu_++h.c.
\end{align}

Writing the Hamiltonian in each case as $H=(\delta/2)\nu_z+A\nu_++A^*\nu_-$ we can write the capacitance as 
\begin{align}
    C=\partial^2 E/\partial\delta^2= \frac{2|A|^2}{(\delta^2 +4|A|^2)^{3/2}},\label{eq:CApp}
\end{align}
where the amplitude for the even and odd case is 
\begin{align}
    |A|^2&=|t_R e^{i\pi\phi}(1-i\zeta_R)+t_L(-i+\zeta_L)|^2\textrm{ for even parity}\nonumber\\
    &=|t_R e^{i\pi\phi}(1+i\zeta_R)+t_L(i+\zeta_L)|^2\textrm{ for odd parity}.\label{eq:AApp}
\end{align}
The $\phi-$dependence of $|A|^2$ in both cases can be written in terms of three parameters $P,Q$ and $R$ as $|A|^2=P+Q\cos{\pi\phi}+R\sin{\pi\phi}$.
In principle, Eqn.~\ref{eq:CApp} together with this parametrization allows a fit for $\delta, P,Q$ and $R$ from the flux dependence of the experimentally measured capacitance $C(\phi)$ over one period. 
 The parameters $P,Q,R$ for each parity sector
 combine to form six parameters, which can be 
 combined using Eq.~\ref{eq:AApp} to determine $t_{L,R}$ and the real and imaginary parts of $\zeta_{L,R}$. The central obstruction to performing this in the current experimental set-up~\cite{aghaee2024interferometric} is the variation of the oscillation between different $hc/e$ periods, which indicates a magnetic-field induced background that must be removed.

Note that the MZM case with $\zeta_{L,R}=0$ 
is characterized by $P$ and $Q$ remaining unchanged between the two parity sectors while $R$ 
changes sign. This corresponds to a $hc/2e$ shift of $\phi$, which is the essential signature expected for a topological phase.
On the other hand, MZM also requires that $\delta$ is set entirely by the QD voltage off-set $V_{QD}$, whose estimation is subject to 
thermal noise. 
Thus, the extent to which the capacitance oscillations show an $hc/2e$ parity shift can bound the spurious coupling of the MZMs though it is 
not a sensitive test of the vanishing of the MZM splitting energy.

\section{Rabi oscillations}\label{S3}
To characterize the coherence of a nanowire-based qubit, we propose using 
Rabi oscillations, in the two Majorana wire set-up coupled by tunneling shown in Fig.~\ref{FIG1}. The Hamiltonian of the system is 
\begin{align}
    &H_{tot}=H_{L}+H_R\nonumber\\
    &+\lambda(\tau) [(t+i t_{SO}\sigma_y)\tau_z\ket{r=N,L}\bra{r=1,R}+h.c],\label{eq:HRabi}
\end{align}
where $H_{L,R}$ are given by Eq.~\ref{eq:HParity} except that the sites are restricted to $r\neq 0$ together with $\lambda_{QD}=0$ to 
remove the QD and $\lambda(\tau)$ is the dimensionless tunneling strength between the wires. 
Note that this set-up differs fundamentally from previous
proposals for studying coherence in nanowire qubits~\cite{Mishmash2020Dephasing},
which requires inter-wire parity measurements. 
 Instead, the set-up in Fig.~\ref{FIG1}, similar to previous experiments~\cite{vanZanten2020Photon}, is based on intra-wire parity measurements together with inter-wire tunneling
that is set to a non-zero value $\lambda(\tau>0)=\lambda$ after initializing with  $\lambda(\tau=0)=0$. 

In this set-up for Rabi oscillations one would initialize the system at $\lambda(\tau=0)=0$ where each nanowire has a single low energy quasiparticle whose occupation is determined by the initial parity and then increase the tunneling to $\lambda(\tau>0)=\lambda$ to generate Rabi oscillations for the quasiparticle to be in the left $\ket{L}$ or right wire $\ket{R}$.
This initializes the system in a superposition of the four lowest energy eigenstates of the final BdG Hamiltonian. The four eigenstates can be classified into groups of two based on the total fermion parity sector. 
Restricting the dynamics to the lowest two positive energy state (one from each nanowire) and considering a specific total Fermion parity sector allows one to map the dynamics using the Bloch-Redfield master equation of a standard two-level spin system~\cite{campaioli2023tutorial} that is characterized by a Rabi oscillation frequency $\Omega$, relaxation time $T_1$, dephasing time $T_2$ as well as an amplitude for the Rabi oscillation of the Fermion parity of the left wire (see Appendix~\ref{SecA2} for details). 

Note that the set-up of the Majorana qubit shown in Fig.~\ref{FIG1} and the qubit degree of freedom described above is essentially similar to a charge qubit~\cite{Bruhat2018Circuit,petersson2010quantum,nakamura1999coherent}.
The dominant source of decoherence in this case is charge noise from Coulomb impurities in the environment, which tends to be long-range correlated. Thus, we will approximate this charge noise as an electrostatic potential that is Markovian i.e. uncorrelated in time but uniform across the length of each wire. The baths of each wire are assumed to be uncorrelated with each other. In the BdG formalism, such a noise Hamiltonian corresponds to uncorrelated in time fluctuations of the chemical potential parameter $\mu$ in Eq.~\ref{eq:HParity}(in Appendix~\ref{SecA1}).
The dynamics of the density matrix generated by the Hamiltonian $H_{tot}$ as well as coupling to baths 
is described by a Bloch-Redfield master equation~\cite{campaioli2023tutorial} (see Appendix~\ref{SecA3} for details).
The effect of the charge noise bath in the wires $L$ and $R$ are represented by parameters $\gamma_{L,R}$ in the master equation~\ref{eq:Master} in Appendix~\ref{SecA3}.
The magnitude of the bath coupling parameters are set to $\gamma_{L,R}\sim 80\, mK$ so as to reproduce conventional charge qubit dephasing~\cite{petersson2010quantum}.

In addition to Rabi oscillations, we will also consider the possibility of estimating the coherence time of Majorana qubits~\cite{Mishmash2020Dephasing}  using 
inter-wire parity or capacitance measurements instead of the inter-wire tunneling required for Rabi oscillations. Such inter-wire parity measurements are necessary to perform  topological operations~\cite{Karzig2017Scalable}. The estimate of dephasing rates in this case takes the form of a decoherence time $T_1'$ for the inter-wire fermion parity. Different choices of ends for the inter-wire fermion parity measurement $X$ and $Y$ are expected to generate different superpositions that differ by a phase shift $\delta\phi$ that can also characterize a topological qubit (see  Appendix~\ref{SecA4} for details).

\subsection{Coherence results}
We first consider the coherence properties of a qubit constructed out of a pair of ideal Majorana nanowires whose 
capacitive read-out properties have been described in Fig.~\ref{FIG2}(a-c). We find, in Fig.~\ref{FIG3}(a,b), that the Rabi oscillation frequency $\Omega$ varies linearly with the tunnel coupling parameter $\lambda$ between the wires. The panels $a$ and $b$ differ by the total 
Fermion parity of the wire system. In contrast, the amplitude of the Rabi oscillation, which is essentially perfect (i.e. equals $0.5$) 
is independent of $\lambda$. The relaxation time $T_1$ and the dephasing time $T_2$ are, as expected, also essentially infinite as seen in these panels. The 
inter-wire parity decoherence rate $(T_1')^{-1}$ is also quite small on this scale.
In contrast, the behavior of the Rabi oscillations for the ABS case corresponding to Fig.~\ref{FIG2}(d-f), which is
shown in Fig.~\ref{FIG3}(c,d), is quite different and shows a Rabi oscillation frequency that varies relatively weakly with
tunnel coupling $\lambda$. Instead the amplitude of the Rabi oscillation is what vanishes as $\lambda$ decreases. Additionally, 
the Rabi oscillation amplitude is strongly suppressed in the even fermion parity sector. This suggests 
that the Rabi oscillation originates from a complicated interplay between inter-wire states and intra-wire ABSs. 
Surprisingly (though consistent with suggestions from prior work~\cite{Vuik2016Effects}), we find that 
the relaxation and dephasing times $T_{1,2}$ and $T_1'$ are still extremely long for the ABS based qubits, though the phase shift $\delta\phi$ is different from the topological value. This clearly implies that the superconductivity
by itself not just  topology can play a strong role in screening the charge noise that generates dephasing in non-superconducting charge 
qubits, which should be readily observable experimentally. 

\begin{figure}[t]
\begin{center}
\includegraphics[width=0.46\textwidth]{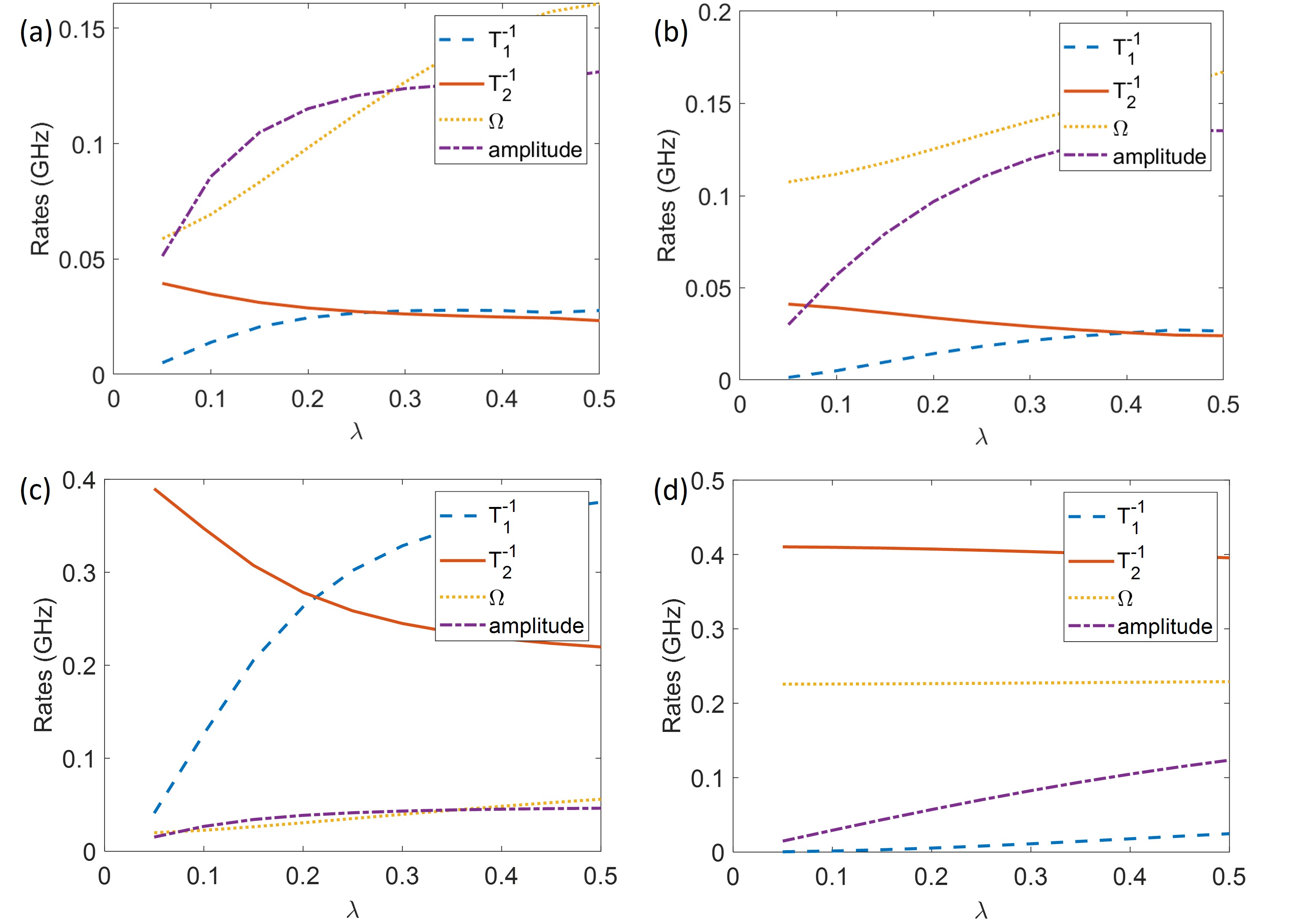}
\end{center}
\vspace{-3mm}
\caption{Rabi oscillation parameters relaxation rate $T_1^{-1}$, dephasing rate $T_2^{-1}$, Rabi-oscillation frequency $\Omega$ and amplitude (dimensionless with scale of $0.3$ for the upper panels, $0.1$  for (c) and $1.1$ for (d)) of Fermion parity oscillation as a function of 
inter-wire coupling $\lambda$. The left panels (right panels) refer to the case for even (odd) total Fermion parity respectively of the system shown in Fig.~\ref{FIG1}. The top panels (a,b) correspond to parameters of Fig.~\ref{FIG25}(a,b,c) with inter-wire parity Rabi oscillation dephasing rate $(T_1')^{-1}\sim 0.02\,GHz$ while the bottom one corresponds to parameters Fig.~\ref{FIG25}(d,e,f) with $(T_1')^{-1}\sim 0.42\,GHz$. }
\label{FIG35}
\vspace{-1mm}
\end{figure}
Fig.~\ref{FIG35} shows Rabi oscillations in weakly disordered Majorana nanowires in the topological patch regime corresponding to the parameters
in Fig.~\ref{FIG25}. The results in Figs.~\ref{FIG35}(a,b) show that the topological regime (i.e. $T_S<0$) corresponding to Figs.~\ref{FIG25}(a,b) 
show relatively coherent Rabi oscillations with relaxation and dephasing rates $T_{1,2}^{-1}$ that are a fraction of the Rabi oscillation frequency provided the dimensionless tunnel coupling $\lambda$ can be increased to a reasonable fraction. This, of course, is much less coherent compared to the disorder free MZM model shown in Fig.~\ref{FIG3}(a,b). 
The crucial requirement of being in the weak disorder limit becomes apparent from the lack of Rabi oscillations with larger disorder seen in Fig.~\ref{FIG35}(c,d), which shows that the Rabi frequency is a small fraction of both the relaxation and dephasing rate $T_{1,2}^{-1}$. 

\section{Summary and discussion}
To summarize our results, motivated by the recent microwave-based Majorana parity read-out experiments~\cite{aghaee2024interferometric}, we have simulated the results of parity measurement for different phases of the semiconductor nanowire system as well as proposed Rabi oscillations as a path to measuring coherence. We find that while the so-called topological patch regime~\cite{das2023spectral} can produce 
parity dependent capacitance signals possibly qualitatively similar to experiments, it does not produce an exact $hc/2e$ parity shift that a topological MZMs produce. Improvements to the current parity-read-out experiment~\cite{aghaee2024interferometric} that eliminate the background flux dependence as well as being sensitive to the $hc/e$ periodic component of the capacitance signal might help distinguish the MZM from the topological patch regime~\cite{aghaee2024interferometric}. 
The time-averaging in the current experiment~\cite{aghaee2024interferometric} leads to the loss of information on the $hc/e$ periodicity so that the claimed periodicity is $hc/2e$. Furthermore, as noted in Sec.~\ref{S2b}, despite similarities to teleportation and fractional Josephson effect, the parity measurement does not directly constrain the MZM splitting $\delta$ beyond temperature. Instead, $\delta$ is determined by transport measurements~\cite{aghaee2022inas}, specifically the zero-bias peak. In this sense, the present capacitance read-out does not contain all features of teleportation proposals~\cite{Fu2010Electron,Sau2015Proposal,Vijay2016Teleportation} where charging energy of the nanowire plays a crucial role.

The single-site approximation discussed in Sec.~\ref{S2}, which is valid in the limit of small QD tunneling, was crucial for the systematic simulation using the BdG formalism as discussed in Sec.~\ref{S2}. However, it is important to discuss possible limitations of the validity of the conclusions if the tunneling to the QD become large compared to the single-particle level spacing. The approximation focuses on the QD level closest to the Fermi energy and assumes that occupation of other levels can be treated as particle-hole excitations from the QD ground state at a specific number of electrons. Given the energy spacing of the single particle excitations being larger than temperature, one does not expect these excitations to participate as non-virtual excitations at low voltages and frequencies. However, these excitations can become important in a virtual sense in the case where the semiconductor nanowire has a large number of levels in a way similar to the Kondo effect~\cite{beri2012topological}, this is not expected to be relevant in this system after it has been studied using transport~\cite{aghaee2022inas}. Therefore, we do not expect a significant effect of the particle-hole exctiations other than renormalization of the effective tunneling constant that is being used as a tuning parameter to match the magnitude of the capacitance. 

The second objective of our work was to study whether 
 the samples studied in Ref.~\cite{aghaee2024interferometric} 
 could demonstrate quantum coherence of the parity degree of freedom in the recent experiment. For this we study measurement of inter-wire fermion parity~\cite{Mishmash2020Dephasing} as well as Rabi oscillations using two wires. In fact, preliminary results of the inter-wire parity measurement have already been presented by the Microsoft group~\cite{nayakAPS2025}, though their results are not currently publicly released. While the inter-wire parity measurements are a simpler way to estimate the coherence time that does not rely on coherent inter-wire tunneling, the Rabi oscillations are potentially more distinctive given an oscillatory pattern is simpler to distinguish from noise. One of the conclusions of our Rabi oscillation simulations shown in Figs.~\ref{FIG3},~\ref{FIG35} is that the coherence of the Rabi oscillation is comparable in magnitude to the inter-wire parity coherence $T_1'$, which  varies by orders of magnitude between the different scenarios considered in Figs.~\ref{FIG25} and ~\ref{FIG35}.
 The inter-wire coherence $T_1'$, in turn, is proportional to a sum of the intra-wire matrix element of the gate charge noise (see Appendix.~\ref{SecA4} for details) for each wire. This allows us to predict the inter-wire parity coherence time for devices where the two wires are in different phases. 
 Considering Eq.~\ref{eq:interwirecoherence} in Appendix~\ref{SecA4}, we conclude that in the case of devices containing wires from different categories in Figs.~\ref{FIG3},~\ref{FIG35} the dephasing rate can be approximated by the higher decoherence rate wire. Since the Rabi oscillation dephasing rate appears to track the inter-wire decay rate $T_1{'-1}$, we can expect the Rabi oscillation decoherence rate to be dominated by the poorer quality wire in the device as well.
 
 The combination of flux dependence capacitance, coherent tunneling Rabi oscillation as well as inter-wire parity Rabi oscillations can  indicate thus providing a direct qubit diagnostic for Majorana nanowires. The signature of MZM then becomes an operational one i.e. enhancement of coherence by topology. A strong topological phase should manifest as an enhanced robustness to charge noise. Such an observation of orders of magnitude enhancement of the coherence time as calculated in our ideal MZM example (Fig.~\ref{FIG3} would be a very strong indication of a topological superconducting phase.

\begin{acknowledgements}
 JS acknowledges valuable discussions with Mike Manfra, Bela Bauer and using KWANT code from Bas Nijoholt's github (https://github.com/basnijholt/majorana-nanowire-conductance). This work is supported by the Laboratory for Physical Sciences and by the Joint Quantum Institute co-directors fund.
\end{acknowledgements}

\bibliography{references.bib}

\appendix
\section{Details of Majorana nanowire model with quantum well and disorder}\label{SecA1}
THe Hamiltonian for the system is written as:
\begin{align}
&H=\sum_{r\leq N} \{(-2 t-\mu+V_{dis}(r))\tau_z+\tau_x\Delta_0+V_Z\sigma_z\}\ket{r}\bra{r}\nonumber\\
&+[\sum_{0<r<N}(t+i t_{SO}\sigma_y)\tau_z\ket{r+1}\bra{r}\nonumber\\
&+\lambda_{QD}(t+i t_{SO}\sigma_y)(\ket{r=N}-e^{i\pi\phi\tau_z}\ket{r=1})\bra{r=0}+h.c],\label{eq:HParity}
\end{align}
where site $r=0$ represents the QD and $\lambda_{QD}$ is the relative weakening of the tunnel barrier between QD and the nanowire. 
 The Hilbert space for the above Hamiltonian, contains a $4$ dimensional Nambu space at each site $r$, with $\sigma_{x,y,z}(\tau_{x,y,z})$ being Pauli matrices in spin (particle-hole) space. The Rashba spin-orbit coupling with strength $\alpha_R$ is perpendicular to the wire and the magnetic field $B_x$ is applied along the nanowire so as to generate with  Zeeman term of strength $V_z=(g/2) \mu_B B_x$ (with the factor chosen to be $g\mu_B=15\,K/T$ based on recent experiments~\cite{aghaee2022inas}).  Based on recent experimental characterization of high quality Majorana nanowire devices~\cite{aghaee2022inas}, we choose $t_0=253.8\, K$ (corresponding to an effective mass $m^*=0.034$ for InAs and lattice parameter $a=10.1\,nm$ ) $t_{SO}=8.59 K$ (
 corresponding to Rashba coupling $\alpha=8\, meV-nm$), and the superconducting coupling is $\Delta_0=1.5\, K$ corresponding to the low frequency 
 limit of the superconducting self-energy~\cite{Sau2010Robustness}.  The hopping efficiency $\lambda_{QD}$ through the QD is chosen to be $\lambda_{QD}=0.03$ so as to obtain energy splittings in the range of severeal tens of mK.
In the above $V_{dis}(r>0)$ is the disorder potential for the Majorana nanowire which is taken to be Gaussian distributed with a correlation length of $\xi_{dis}\sim 30 nm$ with an RMS amplitude $V_{RMS}$, while $V_{dis}(r=0)=V_{QD}$ is the QD potential 
set by a gate. In cases where an end potential generated ABS is present, the zero-energy end ABSs are generated by adding an end square quantum-well
potential of width $L_{abs}=200\,nm$ and depth $V_{abs}=4.5\,K$ into $V_{dis}(r)$ in Eq.~\ref{eq:HParity}. The profile of the quantum 
well is somewhat arbitrarily chosen based on previous work~\cite{Kells2012Near,Prada2012Transport,Rainis2013Towards,Liu2017Andreev,Vuik2019Reproducing} so as to produce low energy localized ABSs as 
seen in Fig.~\ref{FIG2}(c).

The topological stability $T_S$ used to characterize the topological nature of the wire is closely related to the topological invariant of the Majorana nanowire defined by Kitaev~\cite{Kitaev2001Unpaired}. The topological invariant or Majorana number was defined by Kitaev for the Hamiltonian in Eq.~\ref{eq:HParity}, which has periodic/antiperiodic boundary conditions as the ground state fermion parity of $H(\phi)$. The ground state fermion parity of $H$ can be written as $P(\phi)=\textrm{sign}(\textrm{Pffafian}(\sigma_y\tau_y H(\phi)))$. Since, shifting $\phi$ by 1, the topological invariant 
defined by Kitaev~\cite{Kitaev2001Unpaired} can be written as the ratio of the Fermion parity between periodic and anti-periodic boundary conditions $P_{top}=P(\phi)/P(\phi+1)$. This topological invariant provides no insight into the proximity to the topological transition. The Pffafian in $P$ changes sign when the smallest eigenvalue of $H(\phi=0,1)$ crosses $0$. For this reason we define the topological stability 
\begin{align}
&T_S=P_{top}\textrm{min}(E_g(\phi=0,1))/\textrm{max}(E_g(\phi=0,1)),
\end{align} 
where $E_g(\phi)$ is the gap of the Hamiltonian $H(\phi)$.

\section{Technical details of Rabi oscillation calculation}\label{SecA2}

The Rabi oscillation is observed by measuring 
the Fermion parity, which in the low energy sector is given by  $F_{P,L}=\ket{\psi_{1,L}}\bra{\psi_{1,L}}-\ket{\tilde{\psi}_{1,L}}\bra{\tilde{\psi}_{1,L}}$ of $H_L$. Here $\ket{\psi_{1,L}}$ 
and $\ket{\tilde{\psi}_{1,L}}$ are the smallest positive and negative energy eigenstates of the BdG Hamiltonian $H_L$ at $\lambda=0$. 
Particle-hole symmetry relates these two states so that the occupancy of one state guarantees that the other state is unoccupied.
In this work, we  make the simplifying approximation that following the turn-on of the tunneling for $\tau>0$, only the lowest 
two positive energy states $\phi_{1,2}$ (and their particle-hole conjugates $\tilde{\phi}_{1,2}$) with energies $E_{1,2}$ 
can be occupied in a way that is constrained by the total Fermion parity of the system being even or odd.
Particle-hole symmetry means that the occupancy of the other states $\tilde{\phi}_{1,2}$ is determined by the occupancy 
of their particle-hole conjugate. Therefore, the Rabi oscillation for $\tau>0$ can be considered to either be an oscillation between 
the single particle levels $\phi_1$ and $\phi_2$ for the case of odd total parity and $\phi_2$ and $\tilde{\phi}_1$ for the case where 
the total parity is even. This allows one to describe each parity sector by a two-level system subject to bath coupling operators $S_L=\tau_z \sum_r \ket{r,L}\bra{r,L}$ 
and $S_R=\tau_z \sum_r \ket{r,R}\bra{r,R}$ that represent uniform potential coupling to the left and right wires respectively. 
Assuming that the noise on the two wires is uncorrelated, the Linblad form of the Bloch-Redfield master equation~\cite{campaioli2023tutorial} 
is written as:
\begin{align}
    &\dot{\rho}=-i[H_{tot},\rho]\nonumber\\
    &+\sum_{a=L,R;\omega}\gamma_a (\omega) \left[ S_\alpha(\omega)\rho(\tau) S_\alpha^\dagger(\omega)-\frac{1}{2}\{S_\alpha(\omega)S_\alpha^\dagger(\omega),\rho(\tau)\}\right],\label{eq:Master}
\end{align}
where $S_\alpha(\omega)=\sum_{\omega=\omega_a-\omega_b}\Pi_a S_\alpha \Pi_b$ and $\Pi_a$ are projection into eigenstate  $\ket{\omega_a}$ 
of $H_{tot}$ with energy $\omega_a$. Note that $\rho(t)$ is the density matrix containing only the two single-particle levels of $H_{tot}$
 appropriate for the relevant parity sector.
  We choose the limit of large bath temperature so that $\gamma_\alpha(\omega)/\gamma_\alpha(-\omega)=e^{-\omega/T_{bath}}\approx 1$. 
 The parameters $\gamma_{L,R}$ are noise powers that are set to $\gamma_L=80$ mK and $\gamma_R=84$ mK 
consistent with decoherence rates for non-superconducting charge qubits.
The charge noise is assumed to act uniformly on the wire according to $Q_\alpha=\tau_z \sum_r \ket{r,\alpha}\bra{r,\alpha}$, which in the basis of the Rabi oscillator eigenstates is 
$S_{\alpha,a,b}=\expect{\psi_a|Q_\alpha|\psi_b}$.

The Rabi oscillation in the low-energy case for a fixed Hamiltonian and two low energy state maps explicitly to the spin$-1/2$ (i.e. two level system). In the basis of eigenstates  $\ket{\omega_\pm}$ the Hamiltonian $H_{tot}=\epsilon\sigma_z$ and the projector operators are $\Pi_{\pm}=\frac{1\pm \sigma_z}{2}$.
 The solution $\rho(\tau)$ of Eq.~\ref{eq:Master} is a Hermitean matrix with unit trace and therefore can be written in Bloch form as $\rho(\tau)=\mathbf{1}/2+\mathbf{r} (\tau)\cdot\mathbf{\sigma}$, where $\mathbf{r}(\tau)$ is a real $3-$ vector. The Bloch-Redfield equations can be written as $d\mathbf{r}/d\tau=M \mathbf{r}(\tau)$, where $M$ is a real $3\times 3$ matrix. One of the eigenvalues of $M$ is purely real $T_1^{-1}$, while the other two are a complex conjugate pair $T_2^{-1}+i\Omega$. The corresponding eigenvector $r_0$ of $M$ defines the Rabi oscillation of the observable $O$ given by 
\begin{align}
    &\textrm{amplitude}=|Tr[\mathbf{r}_0\cdot\mathbf{\sigma}\tilde{O}]|,
\end{align}
where $\tilde{O}$ defined by the matrix elements $\tilde{O}_{ab}=N\expect{\omega_a|O|\omega_b}$ are the matrix elements of 
the observable $O$ in the low energy subspace and $N$ is a normalization factor chosen so that $Tr[\tilde{O}^2]=1$.
The operator $O$, in these 
nanowire devices, which appear in the Hamiltonian as a modulated tunneling term $\lambda(\tau)O$ can dominate the Hamiltonian in the low-energy sector. In this case, the measurement of the responses $\partial^2 H/\partial\lambda^2\propto O$ 
constitutes a measurement of the operator $O$. For the Rabi oscillation presented in the paper, $O$ is the end-end tunneling in a single wire i.e. 
\begin{align}
    O_Z=\tau_z[e^{i\phi\tau_z}\ket{r=0,L}\bra{r=N,L}+h.c]
\end{align}
applied to the decoupled wires, where $\phi_Z$ is the magnetic flux-induced phase in the tunneling.
In the limit that the wire has a single (pair if one looks at the BdG spectrum) low energy state in the wire, the fermion parity $F_P$ (or equivalently the occupation of this state) is the only degree of freedom so that the operator $O_Z=\bar{O} F_P$ when projected on to the low energy space becomes proportional to the Fermion parity (apart from a shift) discussed in the main text.  

\begin{figure}[t]
\begin{center}
\includegraphics[width=0.65\textwidth]{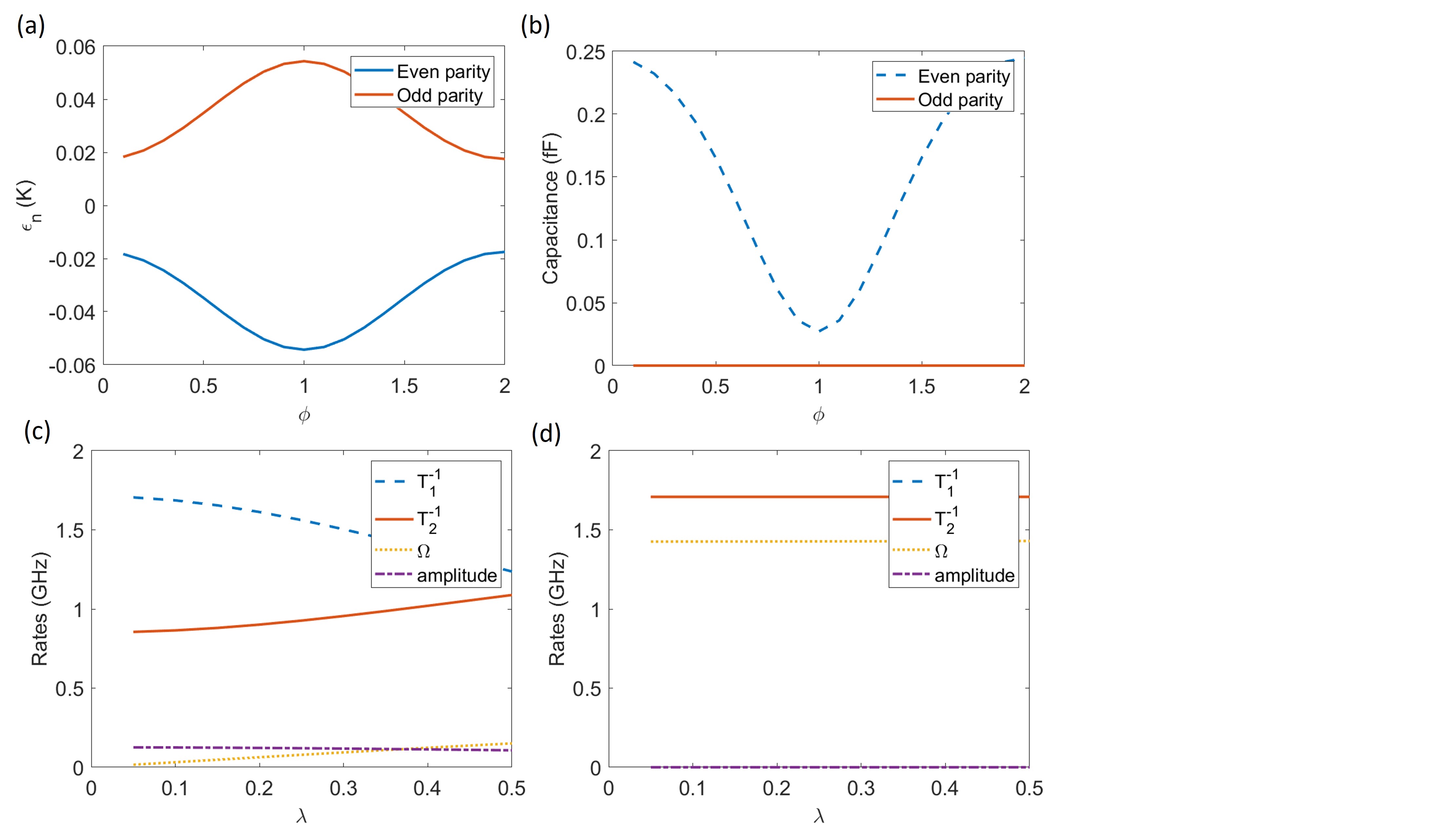}
\end{center}
\vspace{-3mm}
\caption{(a) Energy in K and (b) capacitance for the occupied and unoccupied state as a function of magnetic flux $\phi$ (in units where $hc/e=2$) for an $L=500\,nm$  non-superconducting (i.e. $\Delta_0=0$) nanowire in the topological patch regime~\cite{das2023spectral} at $\mu=-1.62\,K$ and $B_x=0.28\, T$, $V_{QD}=0.0\,K$, $V_{RMS}=0\, K$. The wave-functions for this case shows a single low-energy bound state with a topological stability (defined below Eq.~\ref{eq:HParity}) $T_S=-0.06$. (c,d) Rabi oscillation parameters relaxation rate $T_1^{-1}$, dephasing rate $T_2^{-1}$, Rabi-oscillation frequency $\Omega$ and amplitude (dimensionless with scale of $0.2$ for the upper panels and $14$ for the lower panels) of Fermion parity oscillation as a function of 
inter-wire coupling $\lambda$. Panels (c) and (d) refer to Rabi oscillations in the even and odd total Fermion parity respectively of the system.  }
\label{FIG5}
\vspace{-1mm}
\end{figure}

\section{Rabi oscillation calibration}\label{SecA3}
We calibrate the noise parameters for the master equation by calculating the Rabi oscillations in a nanowire configuration 
similar to the set-up discussed in the main text, other than that superconductivity is removed. Without superconductivity 
the Fermion parity qubit simply because a conventional charge qubit~\cite{petersson2010quantum}.
The result of this non-superconducting case that is used to 
set the value of $\gamma_{L,R}$ is shown in Fig.~\ref{FIG5}.

\section{Inter-wire parity Rabi oscillations}\label{SecA4}
An alternative way to test coherence is to consider the case where 
the tunneling introduced is inter-wire so that 
\begin{align}
    &O_X=\tau_z[e^{i\phi\tau_z}\ket{r=N,L}\bra{r=0,R}+h.c]\\
    &O_Y=\tau_z[e^{i\phi\tau_z}\ket{r=0,L}\bra{r=0,R}+h.c].
\end{align}
Again, in the case where there is a single low energy state in each wire, $O_{X,Y}$ flip the occupancy of the state in each wire and therefore anti-commute with the Fermion parity of the left wire $F_P$. This means that $O_{X,Y}$ also anticommute with $O_Z$.  
This is part of the consistency requirement for an ideal Majorana representation where $O_Z=i\cos{\phi}\gamma_{LL}\gamma_{R,L}$,  $O_X=i\cos{\phi}\gamma_{LL}\gamma_{LR}$ and $O_Y=i\cos{\phi}\gamma_{RL}\gamma_{RR}$ would form three anti-commuting Pauli matrices. In the general case of a single low-energy level in each wire, which encompasses the patch regime as well as the single ABs regime, the anti-commutation between $O_X$ and $O_Y$ is 
not obvious.

To understand the states of the qubit at fixed fermion parity we define the 
$O_Z$ (or $z$-basis) of the qubit as a fermion being on the left nanowire $\ket{L}$ or on the right-nanowire $\ket{R}$. The even or odd total fermion parity is then simply about choosing a reference ground state with even or odd fermion parity. The two states $\ket{L,R}$ are the two eigenstates of $O_Z$ or 
equivalently $F_{P,L}$ (i.e. fermion parity of the left wire). Measurement of the inter-wire Fermion parity $O_{X,Y}$, which anticommutes with $F_{P,L}$ then leads to a superposition state  $\ket{L}\pm e^{\pm i\phi_{X,Y}}\ket{R}$ where $\phi_X$ and $\phi_Y$ are phases associated with the measurement $O_X,O_Y$. A topological qubit would result in a phase-shift $\delta\phi=|\phi_X-\phi_Y|-\pi/2\sim 0$ between the measurements $O_X$ and $O_Y$ that would be necessary for the implementation of Pauli gates in the Clifford algebra~\cite{Karzig2017Scalable}. The calculation of such a phase shift, which can be estimated from measuring the correlation between $X$ and $Y$ on the qubit, provides another comparison to a topological qubit.
We find, however, that all the cases we studied except the ABS wire (i.e. Fig.~\ref{FIG2}(d,e,f)) satisfied this topological criterion.
Experimentally, testing this phase shift condition would amount to testing the orthogonality of $O_X$ and $O_Y$ i.e. whether an $O_X$ eigenstate is a completely mixed $O_Y$ state.

Measurement of inter-wire tunneling allows a few more tests.
Specifically, in this case, one can obtain Rabi oscillations even without
the interwire tunneling.
 The  operators $S_\alpha$ are diagonal in eigenstates, which are localized to $a,b=L,R$.  
Therefore $S_\alpha(\omega)$ is only non-zero at $\omega=0$ and are proportional to projection matrices i.e. $S_\alpha(0)=\Pi_{\alpha}D_\alpha$, where $D_\alpha=S_{\alpha,\alpha\alpha}=\expect{\omega_\alpha|\tau_z|\omega_\alpha}$.
The diagonal elements of $\rho$ then become conserved and the master equation reduces to pure dephasing
\begin{align}
    \dot{\rho}_{12}= -[i(\omega_L-\omega_R)-\frac{(\gamma_L D_L^2+\gamma_R D_R^2)}{2}] \rho_{12}.\label{eq:interwirecoherence}
\end{align}
The dephasing rate in this case is simply $(T_1')^{-1}=(\gamma_L D_L^2+\gamma_R D_R^2)/2$, 
which essentially corresponds to the response of the energy level to charge noise.


\end{document}